\documentclass[journal]{IEEEtran}
\usepackage{amsmath,amsfonts}
\usepackage{algorithmic}
\usepackage{algorithm}
\usepackage{array}
\usepackage[caption=false,font=normalsize,labelfont=sf,textfont=sf]{subfig}
\usepackage{textcomp}
\usepackage{stfloats}
\usepackage{url}
\usepackage{verbatim}
\usepackage{graphicx}
\usepackage{cite}
\usepackage{booktabs}
\usepackage{multirow}
\usepackage{makecell}  
\usepackage{tabularx}
\usepackage[table]{xcolor}
\definecolor{blue}{RGB}{0,112,192}
\usepackage[colorlinks=true,
            linkcolor=blue,
            citecolor=blue,
            urlcolor=blue]{hyperref}
\usepackage{siunitx}
\begin{document}

\title{Profit-Oriented Planning and Multi-Market Operation Model for Hybrid Energy Storage Systems}

\author{Lizhong Zhang,~\IEEEmembership{Student Member,~IEEE}, Junqi Liu, Jianxiao Wang,~\IEEEmembership{Senior Member,~IEEE}, Lei Zhu, ~\IEEEmembership{Member,~IEEE} 

\thanks{Manuscript received May, 2026. (\textit{Corresponding author: Lei Zhu})}
\thanks{
Lizhong Zhang is with School of Economics and Management, Beihang University, Beijing 100191, China (e-mail: zhanglizhong@buaa.edu.cn).

Junqi Liu is with School of Systems Science, Beijing Normal University 100875, Beijing, P. R. China (e-mail: jqliu@bnu.edu.cn).

Jianxiao Wang is with National Engineering Laboratory for Big Data Analysis and Applications, Peking University, Beijing 100871, China (e-mail: wang-jx@pku.edu.cn)

Lei Zhu is with National Safety and Emergency Management, Beijing Normal University 100875, Beijing, P. R. China (e-mail: lions85509050@gmail.com; leizhu@bnu.edu.cn)}}



\maketitle

\begin{abstract}
The increasing penetration of renewable energy necessitates improved power system flexibility, driving the deployment of independent energy storage operators (ESOs). Existing research extensively investigates capacity sizing for price-taker storage systems or the operational coordination of aggregated distributed resources, lacking the joint optimization of capacity planning and multi-market bidding for a price-maker ESO with hybrid energy storage system (HESS) that preserves the technological heterogeneity of the integrated components. We propose a bi-level optimization framework to jointly optimize profit-oriented decisions on capacity and multi-market operation. The upper-level problem determines the optimal capacities of two heterogeneous storage systems while coordinating their bidding across day-ahead joint energy-reserve and real-time balancing markets. The lower-level problems represent market clearing of the system operator (SO). The model is reformulated into a mixed-integer linear program and solved with a Benders' decomposition algorithm. Results demonstrate that the ESO can allocate capacity between energy arbitrage and reserve provision strategically. The system with the high power-to-capacity ratio is used to capture arbitrage profits while the system with low power-to-capacity ratio is used to specialize in reserve markets. There can be internal power transfer between storage systems if there exist grid access constraints. The framework provides differentiated bidding strategies and market participation flexibility for HESS to enhance overall profitability.
\end{abstract}


\centerline{NOMENCLATURE}
\fontsize{10pt}{11pt}\selectfont
\noindent \textit{A. Indices and Sets}

\noindent
\begin{tabularx}{\columnwidth}{@{}l X@{}}
$g$& \quad Index of natural gas generator set $\mathcal{G}$. \\
$j$& \quad Index of typical day set $\mathcal{J}$. \\
$k$& \quad Index of Benders' iteration. \\
$s$& \quad Index of the storage system set $\mathcal{S}=\{s_1, s_2\}$.\\
$t$& \quad Index of the period set $\mathcal{T}=\{1,2,\ldots,T\}$. \\
$w$& \quad Index of wind power generator set $\mathcal{W}$.\\
\end{tabularx}

\vspace{6pt}
\noindent \textit{B. Parameters}

\noindent
\begin{tabularx}{\columnwidth}{@{}l X@{}}
$c_g$& Marginal generation cost of gas generator $g$ (\$/MWh).\\
$c_s$& Marginal charging/discharging cost of storage $s$ (\$/MWh).\\
$C_{s}$& Unit investment cost of storage $s$ (\$/MWh/per day).\\
\end{tabularx}
\vspace{6pt}

\noindent
\begin{tabularx}{\columnwidth}{@{}l X@{}}
$E_s^{max}$& Maximum investment limit of the rated capacity of energy storage $s$ (MWh).\\

$Q_{t,j}^{D}$& Electricity demand at time $t$ of typical day $j$ (MW).\\

$Q_{t,j}^{RU/RD, req}$& System requirement for up/down-reserve at time $t$ of typical day $j$ (MW).\\

$Q^{trans,max}$& Maximum internal power transfer capacity between storage $s_1$ and $s_2$ (MW).\\

$Q_{w,t,j}^{min/max}$& Forecast minimum/maximum wind power generator $w$ at time $t$ of a typical day $j$ (MW).\\

$Q_g^{min/max}$& Minimum/maximum power output of gas generator $g$ (MW).\\

$r_s$& Power-to-capacity ratio of storage $s$.\\

$R_g^{up/down}$& Maximum ramp-up/ramp-down rate of gas generator $g$ (MW/h).\\

$p_{w/g,t,j}^{E,DA}$& Offer price for wind generator $w$/gas generator $g$ at time $t$ of typical day $j$ in the day-ahead market.\\

$q_{w/g,t,j}^{E,DA}$& Offer quantity for wind generator $w$/gas generator $g$ at time $t$ of typical day $j$ in the day-ahead market.\\

$p_{g,t,j}^{E,RT}$& Offer price for energy from gas generator $g$ at time $t$ of typical day $j$ in the real-time market.\\

$q_{g,t,j}^{E,RT}$& Offer quantity for energy from gas generator $g$ at time $t$ of typical day $j$ in the real-time market.\\

$p_{g,t,j}^{RU/RD,DA}$& Offer price for up/down-reserve from gas generator $g$ at time $t$ of typical day $j$ in the day-ahead market.\\

$q_{g,t,j}^{RU/RD,DA}$& Offer quantity for up/down-reserve from gas generator $g$ at time $t$ of typical day $j$ in the day-ahead market.\\

$p^{E/R,DA}$& Maximum price limit for the energy/reserve in the day-ahead market. \\

$p^{E,RT}$& Maximum price limit for the real-time market. \\

$\gamma_j$& Weight of a typical day $j$.\\

$\theta_{min/max}$& Minimum/maximum rate of state of charge (SOC) for storage system $s$ (\%).\\

$\theta_{ini}$& Initial SOC ratio of storage $s$.\\

$\eta_s^{ch/dch}$& Charging/discharging efficiency of storage system.\\
\end{tabularx}

\vspace{6pt}
\noindent \textit{C. Upper-level variables offers of ESO}

\noindent
\begin{tabularx}{\columnwidth}{@{}l X@{}}
$E_{s}$& Rated capacity of storage $s$ (MWh).\\

$E_{s,t,j}$& SOC of storage $s$ at the end of time $t$ of typical day $j$ (MWh).\\

$p_{s,t,j}^{ch/dch,DA}$& Bid/offer prices for charging/discharging of storage $s$ at time $t$ of typical day $j$ in the day-ahead market (\$/MWh).\\
\end{tabularx}

\noindent
\begin{tabularx}{\columnwidth}{@{}l X@{}}
$p_{s,t,j}^{ch/dch,RT}$& Bid/offer prices for charging/discharging of storage $s$ at time $t$ of typical day $j$ in the real-time market (\$/MWh).\\

$p_{s,t,j}^{RU/RD,DA}$& Offer prices for up/down-reserve of storage $s$  at time $t$ of typical day $j$ in the day-ahead market (\$/MWh).\\

$q_{s,t,j}^{ch/dch,DA}$&  Bid/offer quantity for charging/discharging of storage $s$ to the power grid at time $t$ of typical day $j$ in the day-ahead market (MW).\\

$q_{s,t,j}^{ch/dch,RT}$&  Bid/offer quantity for charging/discharging of storage $s$ to the power grid at time $t$ of typical day $j$ in the real-time market (MW).\\

$q_{s,t,j}^{RU/RD,DA}$& Offer quantity for up/down-reserve of storage $s$ at time $t$ of typical day $j$ in the day-ahead market (MW).\\

$q_{s,s',t,j}^{trans}$& Planned power for transfer from storage $s$ to $s'$ at time $t$ of typical day $j$ (MW).\\

$\bar{q}_{s,s',t,j}^{trans}$& Achieved power for transfer from storage $s$ to $s'$ at time $t$ of typical day $j$ (MW).
\end{tabularx}

\noindent \textit{D. Lower-Level Variables (Day-ahead Market Clearing)}

\noindent
\begin{tabularx}{\columnwidth}{@{}l X@{}}
$\bar{q}_{s,t,j}^{ch/dch,DA}$& Clearing charging/discharging quantities of storage $s$ at time $t$ of typical day $j$ (MW).\\

$\bar{q}_{s,t,j}^{RU/RD,DA}$& Clearing reserve quantities of storage $s$ at time $t$ of typical day $j$ (MW).\\

$\bar{q}_{w/g,t,j}^{E,DA}$& Clearing energy quantities of wind/gas power generators at time $t$ of typical day $j$ (MW).\\

$\bar{q}_{g,t,j}^{RU/RD,DA}$& Clearing up/down-reserve quantities of gas power generators at time $t$ of typical day $j$ (MW).\\

$\lambda_{t,j}^{E/RU/RD}$& Market clearing prices for energy, up-reserve, and down-reserve at time $t$ of typical day $j$ (\$/MWh).
\end{tabularx}

\noindent \textit{E. Lower-Level Variables (Real-time Market Clearing)}

\noindent
\begin{tabularx}{\columnwidth}{@{}l X@{}}
$\bar{q}_{s,t,j}^{ch/dch,RT}$& Clearing charging/discharging quantities of storage $s$ at time $t$ of typical day $j$ in the real-time market (MW).\\

$\bar{q}_{s,t,j}^{RU/RD,RT}$& Deployed up/down-reserve quantities of storage $s$ at time $t$ of typical day $j$ in the real-time market (MW).\\

$\bar{q}_{g,t,j}^{E,RT}$& Clearing energy quantities of gas power generators at time $t$ of typical day $j$ in the real-time market (MW).\\

$\bar{q}_{g,t,j}^{RU/RD,RT}$& Deployed up/down-reserve quantities of gas power generators at time $t$ of typical day $j$ in the real-time market (MW).\\

$\lambda_{t,j}^{RT}$& Market clearing prices in the real-time market at time $t$ of typical day $j$ (\$/MWh).
\end{tabularx}
\vspace{2pt}

\noindent \textit{F. Binary Variables for Storage States}

We define generic binary variables $\omega_{s_i,t,j}^{\text{state}} \in \{0, 1\}$ that equal 1 if storage system $s_i$ ($i \in \{1,2\}$) operates in a designated state at time $t$ of typical day $j$, and 0 otherwise. Let $s_{i'}$ denote the other storage system ($i' \neq i$). The seven possible states for system $s_i$ are defined as follows:

\noindent
\begin{tabularx}{\columnwidth}{@{}l X@{}}
$\omega_{s_i,t,j}^{\text{idle}}$& $s_i$ is idle (no charge or discharge).\\

$\omega_{s_i,t,j}^{\text{cg}}$ / $\omega_{s_i,t,j}^{\text{dg}}$& $s_i$ is charging from / discharging to the grid only.\\

$\omega_{s_i,t,j}^{\text{cs}}$ / $\omega_{s_i,t,j}^{\text{ds}}$& $s_i$ is charging from / discharging to $s_{i'}$ only.\\

$\omega_{s_i,t,j}^{\text{cgs}}$ / $\omega_{s_i,t,j}^{\text{dgs}}$& $s_i$ is charging from / discharging to both the grid and $s_{i'}$ simultaneously.
\end{tabularx}

\section{Introduction}
\IEEEPARstart{T}{he} increasing penetration of variable renewable energy sources requires enhanced flexibility to maintain grid stability and balance electricity supply and demand \cite{kaya2026fifty, levin2023energy}. Energy storage systems (ESSs) offer a feasible solution to address these intermittency issues, improve the economic efficiency of power systems, and facilitate deep decarbonization \cite{davies2019combined}. Recent market developments promote the deployment of independent energy storage facilities managed by storage operator \cite{fares2017impacts}. The primary objective of an energy storage operatior (ESO) is to maximize profits through active market participation \cite{bjorndal2023energy}.

Extensive research evaluates the optimal capacity of energy storage using robust and stochastic optimization \cite{billionnet2016robust, ru2013storage}, or co-optimizes the location and size of storage devices considering transmission constraints \cite{fernandez-blanco2017optimal, qi2015joint}. Other studies optimize the sizing of storage devices to participate in multiple electricity markets based on exogenous dynamic pricing \cite{harsha2015optimal, zhang2024fully}. Many of these investment models assume the storage facility acts as a price-taker \cite{lu2004pumpedstorage, jiang2015optimal}. This assumption neglects the substantial impact of large-scale storage deployment on market clearing prices  \cite{cruise2019control}. To capture this strategic impact, researchers tend to use the price-maker paradigm. Bi-level optimization serves as a standard framework to model the interaction between storage operation and market clearing processes \cite{nasrolahpour2018bilevel, khalilisenobari2022optimal, garciat.2025optimal}. Despite the extensive literature on the strategic operation of single-technology storage systems, few studies investigate the joint optimization of capacity planning and multi-market operation for an ESO acting as a price-maker \cite{nasrolahpour2016strategic}.

Reference \cite{zou2016pool} and \cite{zhang2024fully} evaluate the performance of different storage technologies independently but lack the simultaneous consideration of two heterogeneous storage systems within a unified framework. A hybrid energy storage system (HESS) combines heterogeneous technologies to exploit their complementary characteristics, such as pairing high energy density with high power density. Various studies explore the cooperative operation of distributed resources to access wholesale markets \cite{pudjianto2007virtual, mashhour2011bidding}. Coordinating bids across sequential markets or incorporating degradation costs significantly improves overall profitability \cite{lohndorf2022value, kazemi2017operation}. However, existing models rarely embed capacity planning decisions into a strategic multi-market bidding framework that preserves the technological heterogeneity of the hybrid system \cite{pozo2017basic,ruiz2009pool}. Aggregating different storage systems into a single bidding entity would mask the individual operational advantages of the systems. There is a need to develop system-differentiated bidding strategies that allow operators to submit independent price-quantity offers for each technology per market.

While existing literature addresses the impact of network constraints on the operation of storage \cite{wu2023distributed}, current optimization models assume that all energy transfers must occur directly between the storage unit and the main power grid, which could overlook the operational flexibility provided by internal energy routing. An explicit mathematical formulation of the power transfer mechanism among heterogeneous storage systems is required to enable coordinated internal energy management \cite{kemp2023interactions}. This internal coordination provides a critical buffer to sustain market participation and profitability when grid exchange capacities are restricted.

We propose a bi-level optimization framework for an ESO’s decisions of HESS across the day-ahead joint energy and reserve market and the real-time balancing market. The decisions include capacity planning, i.e., storage investment, and operations, i.e., offering and bidding decisions. The upper-level maximizes profit via capacity sizing and bidding, while the lower-level captures market clearing by the system operator (SO). Reformulated as a single-level mathematical programs with equilibrium constraints (MPEC) and solved via Benders' decomposition, the model determines the ESO’s optimal multi-market strategies. The main contributions are as follows.
\begin{enumerate}
\item{We propose a bi-level optimization framework for a price-maker HESS that co-optimizes capacity planning (comprising two heterogeneous storage systems) and participation in both the day-ahead joint energy and reserve market and the real-time balancing market.}
\item{We develop system-differentiated bidding strategies for heterogeneous storage systems by allowing the ESO to submit independent price-quantity offers for each system per market, which exploits the complementary characteristics of diversified storage technologies and tailors market participation accordingly.} 
\item{We explicitly model internal power transmission within the HESS by formulating the power transfer mechanism among heterogeneous storage systems to enable coordinated internal energy management, highlighting the value of internal coordination in sustaining market participation and profitability under limited grid access.}
\end{enumerate}

\section{Framework of the Bi-level Model}
\begin{figure*}[!t]
    \centering
    \includegraphics[width=1\linewidth]{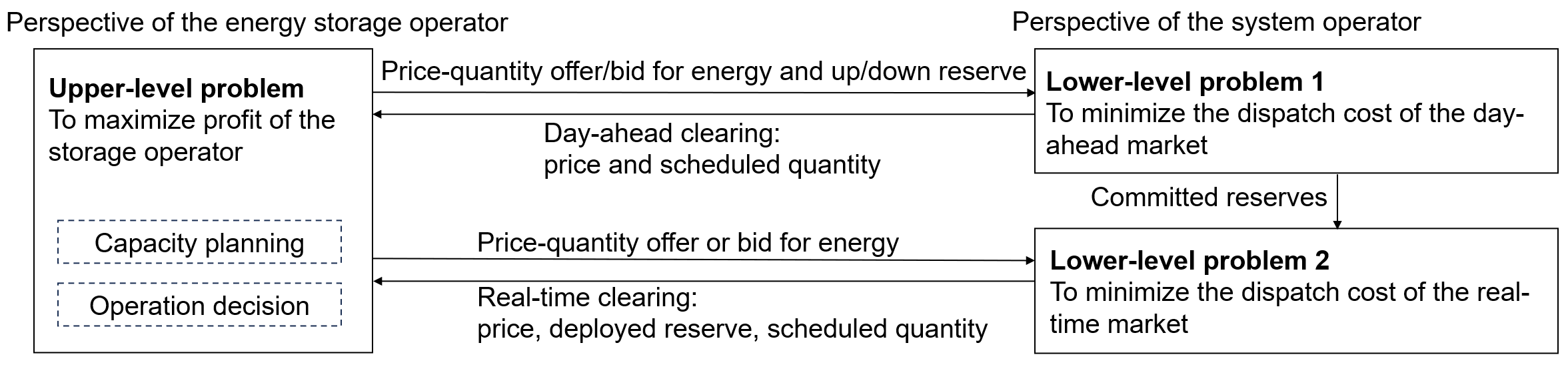}
    \caption{Bi-level model for capacity planning and operational decisions of ESO.}
    \label{fig:bilevel}
\end{figure*}

The bi-level optimization framework shown in Figure\ref{fig:bilevel} is constructed to model the profit-oriented behavior of an independent ESO managing a HESS within the electricity market. In the upper-level problem (ULP), the ESO maximizes profit by jointly optimizing capacity and operation. In the capacity planning aspect, it determines the installed capacities of two heterogeneous storage systems ($s_1$ and $s_2$), embedding amortized capital costs. In the operation aspect, it submits price–quantity offers and bids for charging, discharging, and reserve provision across the day-ahead joint energy and reserve market and the real-time balancing market, while coordinating internal power transfers between systems to manage the overall state of charge and enhance arbitrage flexibility. The decisions formulated in the ULP act as parameters that are fed into the lower-level problems. Meanwhile, the ULP anticipates the market clearing prices and schedules that will result from its submitted bids.

Lower-level problems (LLPs) represent the system operator’s (SO) market clearing. Both LLPs are modeled as bid-based cost minimization problems that process offers from the ULP alongside exogenous offers from other market participants (e.g., wind and gas generators). In this operational framework, the SO acts as a rational follower reacting to the ESO's operations. The optimization of LLP 1 and LLP 2 is parameterized by the ESO's upper-level decisions, meaning every operational move by the ESO inherently alters the feasible region and optimal dispatch of the system.

In LLP 1, the SO clears the day-ahead market by co-optimizing energy and up/down-reserve capacities to minimize total system dispatch costs. Upon clearing, LLP 1 returns the day-ahead energy and reserve clearing prices (derived endogenously as dual variables of the balance constraints) and the scheduled quantities back to the ULP. The cleared reserve quantities from LLP 1 are also passed forward as ``committed reserves" to bound the feasible deployment in the real-time market in LLP 2. Following day-ahead clearing, LLP 2 clears the real-time balancing market to restore system equilibrium at the minimum incremental cost. The SO dispatches resources using the deployed reserves and new real-time energy bids submitted by the market participants. Then LLP 2 returns the real-time clearing prices and the final deployed quantities to the ULP.

The architectural dependency between the ULP and LLPs indicates that the ULP for the ESO cannot determine its maximum profit without evaluating the dual variables (prices) generated by the SO's cost-minimization in the LLPs. Furthermore, there exists an explicit mathematical linkage between the two lower-level problems. Real-time reserve deployment decisions in LLP 2 are strictly constrained by the capacities procured in LLP 1. This coupling forces the ESO to apply multi-market strategies in the ULP that consider the strict certainty of day-ahead commitments and the operational flexibility required for real-time adjustments.

\section{Mathematical Formulation for the Bi-level Model}

In this section, we present the mathematical formulation of the proposed bi-level model, which captures the interplay between the long-term planning and the short-term market operation for an ESO with HESS. The interaction between upper-level and lower-level problems enables us to model ESO as a price-maker. In the bi-level framework, ESO recognizes that its bidding behaviors can actively influence the market clearing prices rather than simply accepting them. Mathematically, this price-maker characteristic is formulated by treating the market clearing prices as endogenous variables derived from the dual formulations of the lower-level problems. Consequently, the long-term planning (capacity $E_s$) dictates the physical boundaries of the HESS, which restricts the feasible space for the short-term operation (price-quantity offers).

\subsection{Upper-Level Problem (ULP): ESO Profit Maximization}

The ESO maximizes the joint profit from its two heterogeneous storage systems, $s_1$ and $s_2$, by coordinating their participation across the day-ahead joint energy-reserve market and the real-time balancing market. The upper-level problem is formulated as follows:
\begin{equation}\label{obj-ulp-eso}
\begin{aligned}
\max_{\Xi_{ES}^{up}} \quad \sum_j \gamma_j\Pi_{ESO,j}-\sum_sC_{s}E_{s}
\end{aligned}
\end{equation}
subject to constraints \eqref{eq:unique_s1}--\eqref{ulp-cons-threshold}, where the set of upper-level decision variables is defined as
$\Xi_{ES}^{up}=\{p_{s,t,j}^{ch,DA},p_{s,t,j}^{dch,DA},p_{s,t,j}^{RU,DA},p_{s,t,j}^{RD,DA}, q_{s,t,j}^{ch,DA}, q_{s,t,j}^{dch,DA}, q_{s,t,j}^{RU,DA},\\  q_{s,t,j}^{RD,DA}, q_{s_1,s_2,t,j}^{trans},q_{s_2,s_1,t,j}^{trans},\bar{q}_{s_1,s_2,t,j}^{trans},\bar{q}_{s_2,s_1,t,j}^{trans}, \omega_{s_i,t,j}^{state}, E_{s,t,j},\\E_s, p_{s,t,j}^{dch,RT},p_{s,t,j}^{ch,RT},q_{s,t,j}^{dch,RT},q_{s,t,j}^{ch,RT}\}$

The objective function comprises two components. The first term represents the weighted sum of the operating profits across all typical days:

\begin{equation}\label{obj-ulp-eso-om}
\begin{aligned}
\Pi_{ESO,j}=&\sum_{t} \bigg\{ \sum_{s} \lambda_{t,j}^E (\bar{q}_{s,t,j}^{dch,DA} - \bar{q}_{s,t,j}^{ch,DA})\\
&-\sum_{s}c_s (\bar{q}_{s,t,j}^{dch,DA}+\bar{q}_{s,t,j}^{ch,DA})\\
&+\sum_{s}c_s ( \lambda_{t,j}^{RU} \bar{q}_{s,t,j}^{RU,DA} + \lambda_{t,j}^{RD} \bar{q}_{s,t,j}^{RD,DA})\\
&- (c_{s_1}+c_{s_2})(\bar{q}_{s_1,s_2,t,j}^{trans}+\bar{q}_{s_2,s_1,t,j}^{trans}) \bigg\}\\
&+\sum_{t} \sum_{s}\bigg\{(\lambda_{t,j}^{RT}-c_{s})(\bar{q}_{s,t,j}^{RU,RT}+\bar{q}_{s,t,j}^{dch,RT})\\
&-(\lambda_{t,j}^{RT}+c_{s})(\bar{q}_{s,t,j}^{RD,RT}+\bar{q}_{s,t,j}^{ch,RT})\bigg\}
\end{aligned}
\end{equation}

The operating profit $\Pi_{ESO,j}$ encompasses revenue from three sources: day-ahead energy market arbitrage, day-ahead reserve provision, and real-time balancing market participation. Operating costs include the marginal cost of charging and discharging operations, as well as internal power transfer costs. Within \eqref{obj-ulp-eso-om}, the profit-oriented operation of the ESO is reflected through the bilinear terms (the product of endogenous clearing prices and cleared quantities, such as $\lambda_{t,j}^{E}\bar{q}_{s,t,j}^{dch,DA}$). These terms mathematically represent the price-maker's trade-off: executing high-volume energy arbitrage or reserve provision will actively shift the equilibrium prices, thereby requiring a bidding operation that balances quantity volumes against adverse price impacts. The second term in \eqref{obj-ulp-eso} represents the total amortized investment cost for the two storage systems.

The constraints governing the upper-level problem are detailed as follows.
\begin{equation}
\begin{aligned}
    \omega_{s_1,t,j}^{\text{idle}} + \omega_{s_1,t,j}^{\text{cg}} + \omega_{s_1,t,j}^{\text{dg}} + \omega_{s_1,t,j}^{\text{cs}} \\+ \omega_{s_1,t,j}^{\text{ds}} + \omega_{s_1,t,j}^{\text{cgs}} + \omega_{s_1,t,j}^{\text{dgs}} = 1 \quad \forall t,j\label{eq:unique_s1}
\end{aligned}
\end{equation}
\begin{equation}
\begin{aligned}
    \omega_{s_2,t,j}^{\text{idle}} + \omega_{s_2,t,j}^{\text{cg}} + \omega_{s_2,t,j}^{\text{dg}} + \omega_{s_2,t,j}^{\text{cs}} \\+ \omega_{s_2,t,j}^{\text{ds}} + \omega_{s_2,t,j}^{\text{cgs}} + \omega_{s_2,t,j}^{\text{dgs}} = 1 \quad \forall t,j\label{eq:unique_s2}
\end{aligned}
\end{equation}
\begin{align}
    \omega_{s_1,t,j}^{\text{ds}} + \omega_{s_1,t,j}^{\text{dgs}} = \omega_{s_2,t,j}^{\text{cs}} + \omega_{s_2,t,j}^{\text{cgs}} \quad \forall t,j \label{eq:logic_s1_to_s2} \\
    \omega_{s_2,t,j}^{\text{ds}} + \omega_{s_2,t,j}^{\text{dgs}} = \omega_{s_1,t,j}^{\text{cs}} + \omega_{s_1,t,j}^{\text{cgs}} \quad \forall t,j \label{eq:logic_s2_to_s1}
\end{align}
\begin{equation}
\begin{aligned}
    E_{s_1,t,j} =  \eta_{s_1}^{ch} (\bar{q}_{s_1,t,j}^{ch,DA} +\bar{q}_{s_2,s_1,t,j}^{trans}+\bar{q}_{s_1,t,j}^{RD,RT}+\bar{q}_{s_1,t,j}^{ch,RT})
    \\-\frac{\bar{q}_{s_1,t,j}^{dch,DA}+\bar{q}_{s_1,s_2,t,j}^{trans}+\bar{q}_{s_1,t,j}^{RU,RT}+\bar{q}_{s_1,t,j}^{dch,RT}}{\eta_{s_1}^{dch}}+E_{s_1,t-1,j}  \quad \forall t,j \label{eq:soc_s1}
\end{aligned}
\end{equation}
\begin{equation}
\begin{aligned}
    E_{s_2,t,j} = \eta_{s_2}^{ch} (\bar{q}_{s_2,t,j}^{ch,DA} +\bar{q}_{s_1,s_2,t,j}^{trans}+\bar{q}_{s_2,t,j}^{RD,RT}+\bar{q}_{s_2,t,j}^{ch,RT})
    \\- \frac{\bar{q}_{s_2,t,j}^{dch,DA}+\bar{q}_{s_2,s_1,t,j}^{trans}+\bar{q}_{s_2,t,j}^{RU,RT}+\bar{q}_{s_2,t,j}^{dch,RT}}{\eta_{s_2}^{dch}} +E_{s_2,t-1,j}\quad \forall t,j \label{eq:soc_s2}
\end{aligned}
\end{equation}
\begin{align}
    \theta_{min}E_s \le E_{s,t,j} \le \theta_{max} E_s \quad \forall s,t,j \label{eq:soc_limits}\\
    0 \le E_s \le E_s^{\max} \quad \forall s \label{eq:capacity_limits}\\
    Q^{trans,max}=\min\{r_sE_s : s\in \mathcal{S}\}\label{status-trans}\\
    0 \le q_{s,t,j}^{dch,DA}\le (\omega_{s,t,j}^{\text{dg}}+\omega_{s,t,j}^{\text{dgs}})r_sE_s \quad \forall s,t,j \label{status-dch}\\
    0 \le q_{s,t,j}^{dch,RT}\le (\omega_{s,t,j}^{\text{dg}}+\omega_{s,t,j}^{\text{dgs}})r_sE_s \quad \forall s,t,j\\
    0 \le q_{s,s',t,j}^{trans} \le (\omega_{s,t,j}^{\text{ds}}+\omega_{s,t,j}^{\text{dgs}}) Q^{trans,max} \quad \forall s,t,j\\
    0 \le q_{s,t,j}^{RU,DA}\le (\omega_{s,t,j}^{\text{dg}}+\omega_{s,t,j}^{\text{dgs}})r_sE_s  \quad \forall s,t,j
\end{align}
\begin{equation}
\begin{aligned}
    0 \le q_{s,t,j}^{dch,DA}+q_{s,s',t,j}^{trans}+q_{s,t,j}^{RU,DA}+q_{s,t,j}^{dch,RT}\\
    \le (\omega_{s,t,j}^{\text{dg}}+\omega_{s,t,j}^{\text{ds}}+\omega_{s,t,j}^{\text{dgs}})r_sE_s \quad \forall s,t,j \label{status-dch-end}
\end{aligned}
\end{equation}
\begin{align}
    0 \le q_{s,t,j}^{ch,DA}\le (\omega_{s,t,j}^{\text{cg}}+\omega_{s,t,j}^{\text{cgs}})r_sE_s \quad \forall s,t,j \label{status-ch}\\
    0 \le q_{s,t,j}^{ch,RT}\le (\omega_{s,t,j}^{\text{cg}}+\omega_{s,t,j}^{\text{cgs}})r_sE_s \quad \forall s,t,j\\
    0 \le q_{s',s,t,j}^{trans} \le (\omega_{s,t,j}^{\text{cs}}+\omega_{s,t,j}^{\text{cgs}}) Q^{trans,max} \quad \forall s,t,j\\
    0 \le q_{s,t,j}^{RD,DA}\le (\omega_{s,t,j}^{\text{cg}}+\omega_{s,t,j}^{\text{cgs}})r_sE_s\quad \forall s,t,j
\end{align}
\begin{equation}
\begin{aligned}
    0 \le q_{s,t,j}^{ch,DA}+q_{s',s,t,j}^{trans}+q_{s,t,j}^{RD,DA}+q_{s,t,j}^{ch,RT}\\
\le(\omega_{s,t,j}^{\text{cg}}+\omega_{s,t,j}^{\text{cs}}+\omega_{s,t,j}^{\text{cgs}})r_sE_s \quad \forall s,t,j\label{status-ch-end}
\end{aligned}
\end{equation}
\begin{align}
    0 \le \bar{q}_{s,s',t,j}^{trans} \le q_{s,s',t,j}^{trans} \quad \forall s,t,j \label{eq:realized-translimit}\\
    0 \le p_{s,t,j}^{ch,DA}, p_{s,t,j}^{dch,DA}\le p^{E,DA}  \quad \forall s,t,j \label{eq-price-1}\\ 
    0 \le p_{s,t,j}^{ch,RT}, p_{s,t,j}^{dch,RT}\le p^{E,RT}  \quad \forall s,t,j\\
        0 \le p_{s,t,j}^{RU,DA}, p_{s,t,j}^{RD,DA}\le p^{R,DA}\quad \forall s,t,j \label{eq-price-3}\\
    E_{s,T,j}\ge\theta_{ini} E_s \quad \forall s,j \label{ulp-cons-threshold}
\end{align}

Constraint \eqref{eq:unique_s1} and \eqref{eq:unique_s2} restrict that at each time step, each storage system can operate in exactly one of seven distinct states. Constraint \eqref{eq:logic_s1_to_s2} and \eqref{eq:logic_s2_to_s1} enforce the logical consistency of internal power transfer. Power can flow from one system to another only if the recipient system is simultaneously in a charging state while the source system is in a discharging state.  Constraint \eqref{eq:soc_s1} and \eqref{eq:soc_s2} represent the SOC balance. The evolution of energy stored in each system is governed by the charging and discharging quantities, as well as internal power transfers, while accounting for round-trip efficiency losses.

Constraint \eqref{eq:soc_limits} preserve that the stored energy in each system must remain within specified bounds throughout the optimization horizon to prevent deep discharge and overcharging. Constraint \eqref{eq:capacity_limits} is the capacity planning limit. Constraint \eqref{status-trans} limits the maximum power transfer capacity between systems. Constraint \eqref{status-dch}-\eqref{status-ch-end} model the discharging and charging limit of the storage systems. Constraint \eqref{eq:realized-translimit} ensures that achieved internal power transfer quantities cannot exceed the planned quantities. Constraint \eqref{eq-price-1}-\eqref{eq-price-3} define the price offer non-negativity and upper bounds. Constraint \eqref{ulp-cons-threshold} ensures that the operator maintains adequate energy reserves for the subsequent operational period, promoting long-term financial viability of the HESS.

\subsection{Lower-Level Problem (LLP) 1: Day-ahead Market Clearing}

For each hour $t$ on a typical day $j$, the SO jointly clears the energy and up-/down-reserve capacities to minimize the total settlement cost evaluated at submitted offer prices. This day-ahead co-optimization yields the cleared schedules $\bar{q}_{\cdot,t,j}^{(\cdot),DA}$ and the day-ahead market prices $(\lambda_{t,j}^{E},\lambda_{t,j}^{RU},\lambda_{t,j}^{RD})$, where each price is the shadow value of the corresponding balance or requirement constraint.

The objective function of the SO in the day-ahead market for $t$ on a typical day $j$ is as follows:
\begin{equation}\label{llp-obj}
\begin{aligned}
\min_{\Xi_1^{low}}  \sum_{w} p_{w,t,j}^{E,DA} \bar{q}_{w,t,j}^{E,DA} + \sum_{g} p_{g,t,j}^{E,DA} \bar{q}_{g,t,j}^{E,DA} \\+ \sum_{s}\left(p_{s,t,j}^{dch,DA} \bar{q}_{s,t,j}^{dch,DA} - p_{s,t,j}^{ch,DA} \bar{q}_{s,t,j}^{ch,DA}\right) \\ + \sum_{g} \left(p_{g,t,j}^{RU,DA} \bar{q}_{g,t,j}^{RU,DA} +p_{g,t,j}^{RD,DA} \bar{q}_{g,t,j}^{RD,DA}\right)\\+ \sum_{s}\left(p_{s,t,j}^{RU,DA} \bar{q}_{s,t,j}^{RU,DA} +p_{s,t,j}^{RD,DA} \bar{q}_{s,t,j}^{RD,DA}\right)
\end{aligned}
\end{equation}
subject to the following constraints:
\begin{equation}
\begin{aligned}
    \sum_w \bar{q}_{w,t,j}^{E,DA} + \sum_g \bar{q}_{g,t,j}^{E,DA} \\+ \sum_{s}\left(\bar{q}_{s,t,j}^{dch,DA} - \bar{q}_{s,t,j}^{ch,DA}\right) -Q_{t,j}^D =0\quad : \lambda_{t,j}^E \label{llp1-eb}
\end{aligned}
\end{equation}
\begin{align}
    \sum_{s}\bar{q}_{s,t,j}^{RU,DA} + \sum_g \bar{q}_{g,t,j}^{RU,DA} -Q_{t,j}^{RU, req}=0  \quad &: \lambda_{t,j}^{RU} \label{llp-urr}\\
    \sum_{s}\bar{q}_{s,t,j}^{RD,DA} + \sum_g \bar{q}_{g,t,j}^{RD,DA} -Q_{t,j}^{RD, req}= 0 \quad &: \lambda_{t,j}^{RD} \label{llp-drr}
\end{align}
\begin{align}
    &0\leq \bar{q}_{w,t,j}^{E,DA}\leq q_{w,t,j}^{E,DA} &&:\underline{\mu}_{w,t,j}^E,\bar{\mu}_{w,t,j}^E &&&\forall w\label{llp-cons-wind}\\
    &0\leq \bar{q}_{g,t,j}^{E,DA}\leq q_{g,t,j}^{E,DA} &&:\underline{\mu}_{g,t,j}^E,\bar{\mu}_{g,t,j}^E  &&&\forall g\\
  &0\leq \bar{q}_{s,t,j}^{ch,DA}\leq q_{s,t,j}^{ch,DA} &&:\underline{\mu}_{s,t,j}^{ch},\bar{\mu}_{s,t,j}^{ch}  &&&\forall s\\
    &0\leq \bar{q}_{s,t,j}^{dch,DA}\leq q_{s,t,j}^{dch,DA} &&:\underline{\mu}_{s,t,j}^{dch},\bar{\mu}_{s,t,j}^{dch}  &&&\forall s\\
    &0\leq\bar{q}_{s,t,j}^{RU,DA}\leq q_{s,t,j}^{RU,DA} &&: \underline{\mu}_{s,t,j}^{RU},\bar{\mu}_{s,t,j}^{RU} &&&\forall s\\
   & 0\leq\bar{q}_{s,t,j}^{RD,DA}\leq q_{s,t,j}^{RD,DA} &&: \underline{\mu}_{s,t,j}^{RD},\bar{\mu}_{s,t,j}^{RD} &&&\forall s\\
    &0\leq\bar{q}_{g,t,j}^{RU,DA}\leq q_{g,t,j}^{RU,DA} &&: \underline{\mu}_{g,t,j}^{RU},\bar{\mu}_{g,t,j}^{RU} &&&\forall g\\
    &0\leq\bar{q}_{g,t,j}^{RD,DA}\leq q_{g,t,j}^{RD,DA} &&: \underline{\mu}_{g,t,j}^{RD},\bar{\mu}_{g,t,j}^{RD} &&&\forall g\label{llp-cons-gas-rd}
\end{align}
where the set of lower-level decision variables is $\Xi_1^{low}=\{\bar{q}_{w,t,j}^{E,DA},\bar{q}_{g,t,j}^{E,DA},\bar{q}_{s,t,j}^{dch,DA},\bar{q}_{s,t,j}^{ch,DA}, \bar{q}_{s,t,j}^{RU,DA},\bar{q}_{s,t,j}^{RD,DA},\bar{q}_{g,t,j}^{RU,DA},\\\bar{q}_{g,t,j}^{RD,DA}\}$ for $w\in\mathcal{W}$, $s\in \mathcal{S}$, and $g\in\mathcal{G}$. The notation after the colon in each constraint indicates the corresponding dual variable(s).

The energy balance constraint \eqref{llp1-eb} ensures that the total energy supplied by wind generators, gas generators, and the energy storage system equals the system demand. The clearing price for the energy market, denoted $\lambda_{t,j}^E$, represents the marginal cost of meeting one additional unit of demand. The up- and down-reserve requirement constraints in constraint \eqref{llp-urr} and \eqref{llp-drr} ensure that sufficient reserve capacity is procured to maintain system stability. The corresponding reserve prices, $\lambda_{t,j}^{RU}$ and $\lambda_{t,j}^{RD}$, represent the marginal value of procuring an additional unit of up- or down-reserve capacity. The bounds on cleared quantities in the last set of constraints \eqref{llp-cons-wind}-\eqref{llp-cons-gas-rd} ensure that no market participant clears more than it has offered.

\subsection{Lower-Level Problem (LLP) 2: Real-time Market Clearing}

After day-ahead clearing, uncertainty is realized in real time through wind forecast error, represented by the net deviation term $\Delta Q_{t,j}$ in hour $t$. The SO then clears the real-time balancing market to restore system balance at minimum incremental cost, using two types of resources: the deployment of cleared day-ahead reserves from both gas generators and the energy storage system, and additional real-time energy and charging/discharging offers. Importantly, the real-time reserve deployment decisions are coupled with the day-ahead outcomes through feasibility constraints, which ensure that only previously procured reserve capacities can be activated in real time. Together, LLP 1 and LLP 2 describe a sequential day-ahead to real-time settlement structure with explicit day-ahead to real-time coupling, which is essential for capturing the ESO's multi-market behavior.

The objective function of the SO in the real-time market for $t$ on a typical day $j$ is as follows:
\begin{equation}\label{llp2-obj}
\begin{aligned}
\min_{\Xi_2^{low}} \sum_{g} p_{g,t,j}^{E,DA} \left( \bar{q}_{g,t,j}^{RU,RT}-\bar{q}_{g,t,j}^{RD,RT}\right)\\ +  \sum_s\left(p_{s,t,j}^{dch,DA} \bar{q}_{s,t,j}^{RU,RT} - p_{s,t,j}^{ch,DA} \bar{q}_{s,t,j}^{RD,RT}\right) \\ + \sum_{g} p_{g,t,j}^{E,RT} \bar{q}_{g,t,j}^{E,RT}\\+ \sum_s\left(p_{s,t,j}^{dch,RT} \bar{q}_{s,t,j}^{dch,RT} -p_{s,t,j}^{ch,RT} \bar{q}_{s,t,j}^{ch,RT}\right)
\end{aligned}
\end{equation}
subject to the following constraints:
\begin{equation}
\begin{aligned}
     \sum_s\left(\bar{q}_{s,t,j}^{RU,RT}-\bar{q}_{s,t,j}^{RD,RT}+\bar{q}_{s,t,j}^{dch,RT}-\bar{q}_{s,t,j}^{ch,RT}\right)\\
     +\sum_g\left(\bar{q}_{g,t,j}^{RU,RT}-\bar{q}_{g,t,j}^{RD,RT}+\bar{q}_{g,t,j}^{E,RT}\right)-\Delta Q_{t,j}=0  \quad :\lambda_{t,j}^{RT} \label{llp2-eb}
\end{aligned}
\end{equation}
\begin{align}
    &0\leq\bar{q}_{g,t,j}^{RU,RT}\leq \bar{q}_{g,t,j}^{RU,DA} &&: \underline{\nu}_{g,t,j}^{RU},\bar{\nu}_{g,t,j}^{RU} &&&\forall g \label{llp2-eq-gas}\\
    &0\leq\bar{q}_{g,t,j}^{RD,RT}\leq \bar{q}_{g,t,j}^{RD,DA} &&: \underline{\nu}_{g,t,j}^{RD},\bar{\nu}_{g,t,j}^{RD}&&&\forall g\\
     &0\leq\bar{q}_{s,t,j}^{RU,RT}\leq \bar{q}_{s,t,j}^{RU,DA} &&: \underline{\nu}_{s,t,j}^{RU},\bar{\nu}_{s,t,j}^{RU} &&&\forall s\\
    &0\leq\bar{q}_{s,t,j}^{RD,RT}\leq \bar{q}_{s,t,j}^{RD,DA} &&: \underline{\nu}_{s,t,j}^{RD},\bar{\nu}_{s,t,j}^{RD} &&&\forall s \label{llp2-eq-storage}\\
    &0\leq\bar{q}_{g,t,j}^{E,RT}\leq q_{g,t,j}^{E,RT} &&: \underline{\nu}_{g,t,j}^{RT},\bar{\nu}_{g,t,j}^{RT} &&&\forall g\label{llp2-cons-gas-rt}\\
    &0\leq\bar{q}_{s,t,j}^{dch,RT}\leq q_{s,t,j}^{dch,RT} &&: \underline{\nu}_{s,t,j}^{dch,RT},\bar{\nu}_{s,t,j}^{dch,RT} &&&\forall s\\
    &0\leq\bar{q}_{s,t,j}^{ch,RT}\leq q_{s,t,j}^{ch,RT} &&: \underline{\nu}_{s,t,j}^{ch,RT},\bar{\nu}_{s,t,j}^{ch,RT} &&&\forall s\label{llp2-cons-es-rt}
\end{align}
where the set of lower-level decision variables is $\Xi_2^{low}=\{\bar{q}_{g,t,j}^{RU,RT},\bar{q}_{g,t,j}^{RD,RT},\bar{q}_{s,t,j}^{RU,RT},\bar{q}_{s,t,j}^{RD,RT}, \bar{q}_{g,t,j}^{E,RT},\bar{q}_{s,t,j}^{dch,RT},\bar{q}_{s,t,j}^{ch,RT}\}$ for $s\in \mathcal{S}$ and $g\in\mathcal{G}$. The notation after the colon in each constraint indicates the corresponding dual variable(s).

The real-time energy balance constraint in \eqref{llp2-eb} ensures that the net energy injection from deploying reserves and new real-time offers equals the net load deviation caused by wind forecast errors. The term $\Delta Q_{t,j}=\sum_w\left(\bar{q}_{w,t,j}^{E,DA}-q_{w,t,j}^{real}\right)$ represents the total deviation between the day-ahead cleared wind energy and the realized wind output. The real-time energy price $\lambda_{t,j}^{RT}$ reflects the marginal cost of balancing this deviation. The reserve deployment constraints in constraint \eqref{llp2-eq-gas}--\eqref{llp2-eq-storage} ensure that the deployment of up- and down-reserves in real time cannot exceed the quantities previously cleared in the day-ahead market. Constraint \eqref{llp2-cons-gas-rt}-- \eqref{llp2-cons-es-rt} enforce that cleared quantities do not exceed the submitted offers.

\section{Solution Technique}

Replacing the lower-level problems with their KKT optimality conditions transforms the original bi-level problem into a single-level MPEC. The complementarity conditions can be linearized via a Big-$M$ reformulation using auxiliary binary variables, and the remaining bilinear terms involving dual variables and day-ahead cleared reserve quantities can be  fully linearized through dual theory \cite{nasrolahpour2018bilevel}. The resulting model is cast as a mixed-integer linear program (MILP).

To efficiently solve the resulting large-scale MILP, we employ a Benders' decomposition that separates long-term investment variables (energy capacities $E_s$) from short-term operational decisions across typical days. While the decomposition framework follows \cite{kazempour2012strategica}, our implementation exploits the problem’s block-angular structure by generating distinct optimality cuts for each typical day rather than a single aggregated cut, to prevent information loss. The algorithm decomposes the problem into:

\begin{enumerate}
    \item{A master problem (MP) that determines the capacities $E_s$ to maximize the total expected profit.}
    \item{A set of Auxiliary Problems (APs) (one per typical day $j$) that solve the operation problem with fixed capacities to identify the optimal binary variables.}
    \item{A set of Linear Subproblems (SPs) (one per typical day $j$) that use the optimal solutions from the AP to create a continuous linear programming (LP), providing the sensitivities $\pi_{s,j}^{(k)}$ for Benders' cuts.}
\end{enumerate}

Let $k$ denotes the iteration counter. The MP determines the capacities $E_s$ for the storage systems. It replaces the complex operational SPs with the proxy variable $\alpha_{j}$ representing the minus expected operational profit for typical day $j$, bounded by Benders' cuts:

\begin{equation} \label{eq:master_obj}
    \max_{\Xi_{MP}} \quad Z_{MP}^{(k)} = -\sum_{j \in \mathcal{J}} \alpha_{j}^{(k)} - \sum_{s} C_s E_s^{(k)}
\end{equation}
subject to:
\begin{align}
    & 0 \leq E_s^{(k)} \leq E_s^{max} , \quad \forall s \in \mathcal{S}\\
    & \alpha_{j}^{(k)} \ge \alpha_j^{min}, \quad \forall j \in \mathcal{J} \label{eq:alpha_bound}
\end{align}
\begin{equation}
\begin{aligned}
    \alpha_{j}^{(k)} \ge \gamma_j \left[ -\Pi_{ESO,j}^{\text{linear},(l)} + \sum_{s} \pi_{s,j}^{(l)} (E_s^{(k)} - E_s^{(l)}) \right], \\\quad \forall j \in \mathcal{J},\forall l = 1, \dots, k-1 \label{eq:benders_cut} 
\end{aligned}
\end{equation}
where $\Xi_{MP}=\{E_{s}^{(k)}, \alpha_{j}^{(k)}\}$ is the set of optimization variables. Constraint (\ref{eq:alpha_bound}) imposes a lower bound on $\alpha_j^{(k)}$ to accelerate convergence.
At every iteration, a new constraint is added to (\ref{eq:benders_cut}). It represents the optimality cuts generated from previous iterations.
$E_s^{(l)}$ is the fixed capacity determined at iteration $l$.
$\Pi_{ESO,j}^{(l)}$ is the operational profit of day $j$ calculated in the SP at iteration $l$.
$\pi_{s,j}^{(l)}$ is the sensitivity (dual variable) of the operational profit with respect to capacity $E_s^{(l)}$ for day $j$ at iteration $l$.
Constraint (\ref{eq:benders_cut}) represents the optimality cuts generated from previous iterations.

For a given iteration $k$ and typical day $j$, AP is formulated as a MILP solved under fixed investment capacities ($E_s^{(k)}$) to handle non-convexities and determine the optimal binary variables. The objective is as follows:
\begin{equation}
    \max_{\Xi_{AP}} \quad \gamma_j\Pi_{ESO,j}^{\text{linear}, (k)}
\end{equation}

The feasible region of the AP is defined by the complete single-level MPEC constraints. These encompass the upper-level physical and logical boundaries, such as storage state exclusivity and state-of-charge evolution, alongside the lower-level day-ahead and real-time market clearing balances. To ensure market equilibrium, the AP embeds the lower-level stationarity and complementarity conditions—the latter explicitly linearized via the Big-M method using auxiliary binary variables. Furthermore, McCormick envelopes are introduced to accurately linearize the bilinear terms present in the upper-level objective, specifically the products of endogenous clearing prices and scheduled quantities \cite{kazempour2012strategica}. The solution of AP per day gives the optimal values of the following variables. We use $\Xi_{AP}^{\text{fix}}=\{\omega_{s,t,j}^{state},p_{s,t,j}^{dch,DA},p_{s,t,j}^{ch,DA},p_{s,t,j}^{RU,DA},p_{s,t,j}^{RD,DA},p_{s,t,j}^{dch,RT},p_{s,t,j}^{ch,RT},\\\bar{\mu}_{s,t,j}^{dch},\bar{\mu}_{s,t,j}^{ch},\bar{\mu}_{s,t,j}^{RU},\bar{\mu}_{s,t,j}^{RD},\bar{\nu}_{s,t,j}^{dch,RT},\bar{\nu}_{s,t,j}^{ch,RT},\bar{\nu}_{s,t,j}^{RU},\bar{\nu}_{s,t,j}^{RD}\}$ to denote the set of these variables that will be fixed in the SP.

The SP computes the accurate sensitivities (dual variables) $\pi_{s,j}^{(k)}$ needed for the Benders' cuts. It is constructed by fixing $\Xi_{AP}^{\text{fix}}$ found in the AP. This transforms the problem into a continuous linear program.
\begin{equation}
    \max_{\Xi_{SP}} \quad \gamma_j\Pi_{ESO,j}^{\text{linear}, (k)}
\end{equation}
subject to:
\begin{align}
    & E_s = E_s^{(k)} \quad : \pi_{s,j}^{(k)} \quad \forall s \label{eq:fix_invest_SP_1} 
\end{align}

In addition to constraint \eqref{eq:fix_invest_SP_1}, the complete set of constraints includes all continuous operational constraints, market balance equations, and the relaxed stationarity and McCormick conditions. Critically, the SP integrates strong duality equations to enforce strict equivalence between the primal and dual objective functions of the lower-level markets \cite{kazempour2012strategica}.  The execution procedure is summarized in Algorithm \ref{alg:benders}.

\begin{algorithm}[!t]
\caption{Benders' Decomposition for the problem}
\label{alg:benders}
\begin{algorithmic}[1]
\STATE Initialize $k=1$, $Z_{MP}^{(1)}=+\infty$, $E_s^{(1)} = E_s^{\text{initial}}$, tolerance $\epsilon>0$.
\WHILE{True}
    \FOR{each typical day $j \in \mathcal{J}$}
        \STATE Solve AP (MILP) with $E_s = E_s^{(k)}$ $\rightarrow$ store $\Xi_{AP}^{\text{fix}}$.
        \STATE Solve SP (LP) with fixed $\Xi_{AP}^{\text{fix}}$ $\rightarrow$ store $\Pi_{ESO,j}^{\text{linear},(k)}$, $\pi_{s,j}^{(k)}$.
    \ENDFOR
    \STATE Compute $\widetilde{\Pi}^{(k)} = \sum_j \gamma_j \Pi_{ESO,j}^{\text{linear},(k)} - \sum_s C_s E_s^{(k)}$.
    \IF{$|Z_{MP}^{(k)} - \widetilde{\Pi}^{(k)}| \le \epsilon$}
        \RETURN $E_s^* = E_s^{(k)}$.
    \ENDIF
    \STATE Add Benders' cuts; solve MP $\rightarrow$ update $E_s^{(k+1)}$, $Z_{MP}^{(k+1)}$.
    \STATE $k \leftarrow k+1$.
\ENDWHILE
\end{algorithmic}
\end{algorithm}

\section{Numerical results}
\subsection{Parameter Settings}

The optimization framework adopts a 24-hour scheduling horizon across four typical days, each assigned an equal weight of 0.25 to represent amortized expected operation. System demand profiles are proportionally scaled from historical aggregated operational data of the New South Wales region in the Australian national electricity market, with the first days of January, April, July, and October selected to capture seasonal variations. Generation data for the participating wind power units are sourced from actual measurements from the Alberta electric system operator (AESO) in Canada and proportionally scaled to the test system magnitude, with forecast values uniformly generated between 80\% and 140\% of realized wind output to reflect prediction uncertainties. Wind energy price offers are set to zero to ensure priority dispatch.

The thermal generation fleet comprises eight gas generators. These generators submit price and quantity bids across the day-ahead energy, day-ahead reserve, and real-time markets based on their intrinsic physical properties. Detailed parameters for bid quantities and prices across the multiple markets are listed in Table \ref{tab:gas_parameters}. The system requirement for up-reserve and down-reserve employs a $3+5$ reserve policy for setting the regulation targets, which sets the reserve requirements to equal 3 percent of the hourly demand plus 5 percent of the renewable output \cite{xu2017scalable}.


\begin{table*}[t]
\centering
\caption{Operational parameters and bid information of the gas generators}
\label{tab:gas_parameters}
\footnotesize
\begin{tabular}{
    crrrrrr
}
\hline
\multirow{2}{*}{Generator} 
& \multicolumn{2}{c}{Day-ahead energy} 
& \multicolumn{2}{c}{Day-ahead reserve} 
& \multicolumn{2}{c}{Real-time energy} \\
\cmidrule(lr){2-3} \cmidrule(lr){4-5} \cmidrule(lr){6-7}
& {Quantity (MW)} & {Price (\$/MWh)} 
& {Quantity (MW)} & {Price (\$/MWh)} 
& {Quantity (MW)} & {Price (\$/MWh)} \\
\hline
g1 & 240 &  20 & 30 &  10 & 30 &  20 \\
g2 & 240 &  30 & 30 &  15 & 30 &  30 \\
g3 & 240 &  40 & 30 &  20 & 30 &  40 \\
g4 & 240 &  60 & 30 &  30 & 30 &  60 \\
g5 & 240 &  90 & 30 &  45 & 30 &  90 \\
g6 & 240 & 120 & 30 &  60 & 30 & 120 \\
g7 & 320 & 150 & 40 &  75 & 40 & 150 \\
g8 & 320 & 300 & 40 & 150 & 40 & 300 \\
\hline
\end{tabular}
\end{table*}

The corresponding technical and economic parameters characterizing the two candidate storage systems are summarized in Table \ref{tab:storage_parameters} based on \cite{zhang2024fully}. The initial state of charge boundary condition begins at 0.50 with marginal operational costs assumed to be zero. The daily investment costs applied in the objective function reflect a linear amortization of the total annual investment cost per year. 

\begin{table}[!t]
\centering
\caption{Technical and economic parameters of the candidate storage systems}
\label{tab:storage_parameters}
\begin{tabular}{lcc}
\hline
Parameter & Storage 1 ($s_1$) & Storage 2 ($s_2$) \\
\hline
Power-to-capacity ratio & 0.2 & 5.0 \\
Maximum capacity limit (MWh) & 500 & 500 \\
Charging and discharging efficiency & 0.95 & 0.95 \\
Minimum state of charge & 0.05 & 0.05 \\
Maximum state of charge & 0.95 & 0.95 \\
Initial state of charge & 0.50 & 0.50 \\
Total investment cost (\$/MWh/year) & 3571 & 4571 \\
\hline
\end{tabular}
\end{table}

\subsection{Capacity Planning and Market Clearing Results}

\begin{figure}[!t]
    \centering
    \includegraphics[width=1\linewidth]{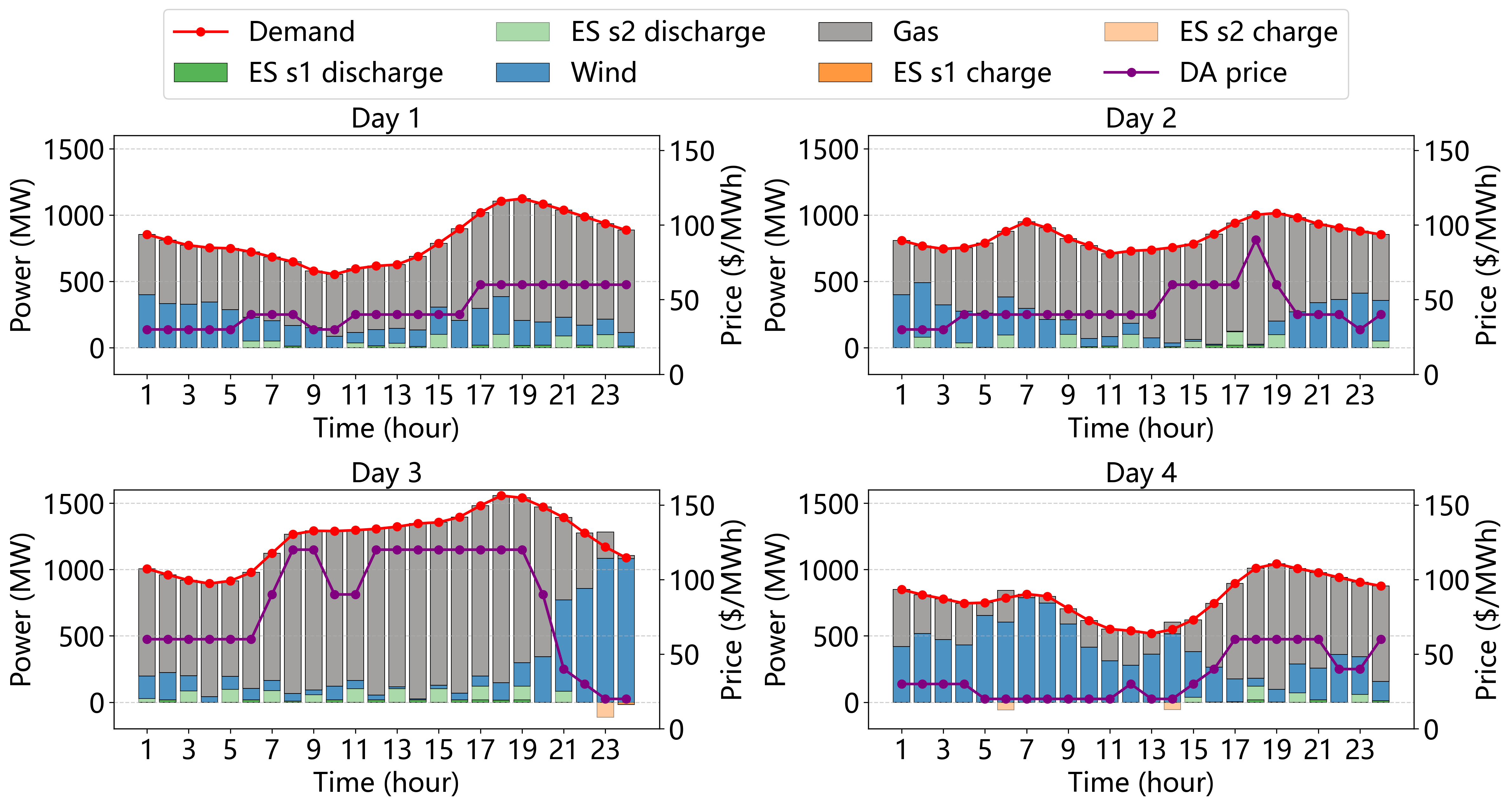}
    \caption{Energy balance of the joint day-ahead market.}
    \label{fig-da-energy}
\end{figure}

The optimal capacities of the two storage systems are 93.89 MWh for $s_1$ with low power-to-capacity ratio and 118.94 MWh for $s_2$ with high power-to-capacity ratio. This mixed investment indicates that a complementary hybrid storage portfolio is profitable to capture diverse revenue streams in the market. Figure \ref{fig-da-energy} illustrates the energy balance within the joint day-ahead market across four typical days. Due to zero offers, wind power is prioritized in the dispatch process. Consequently, high wind power output drives low electricity prices, prompting the ESO to charge. Conversely, the ESO chooses to discharge when prices are high. Rather than acting as a passive price-taker that merely shifts peak load, the ESO strategically coordinates the charging and discharging phases of the two storage systems to maximize arbitrage profits under price volatility.

\begin{figure}[!t]
    \centering
    \includegraphics[width=1\linewidth]{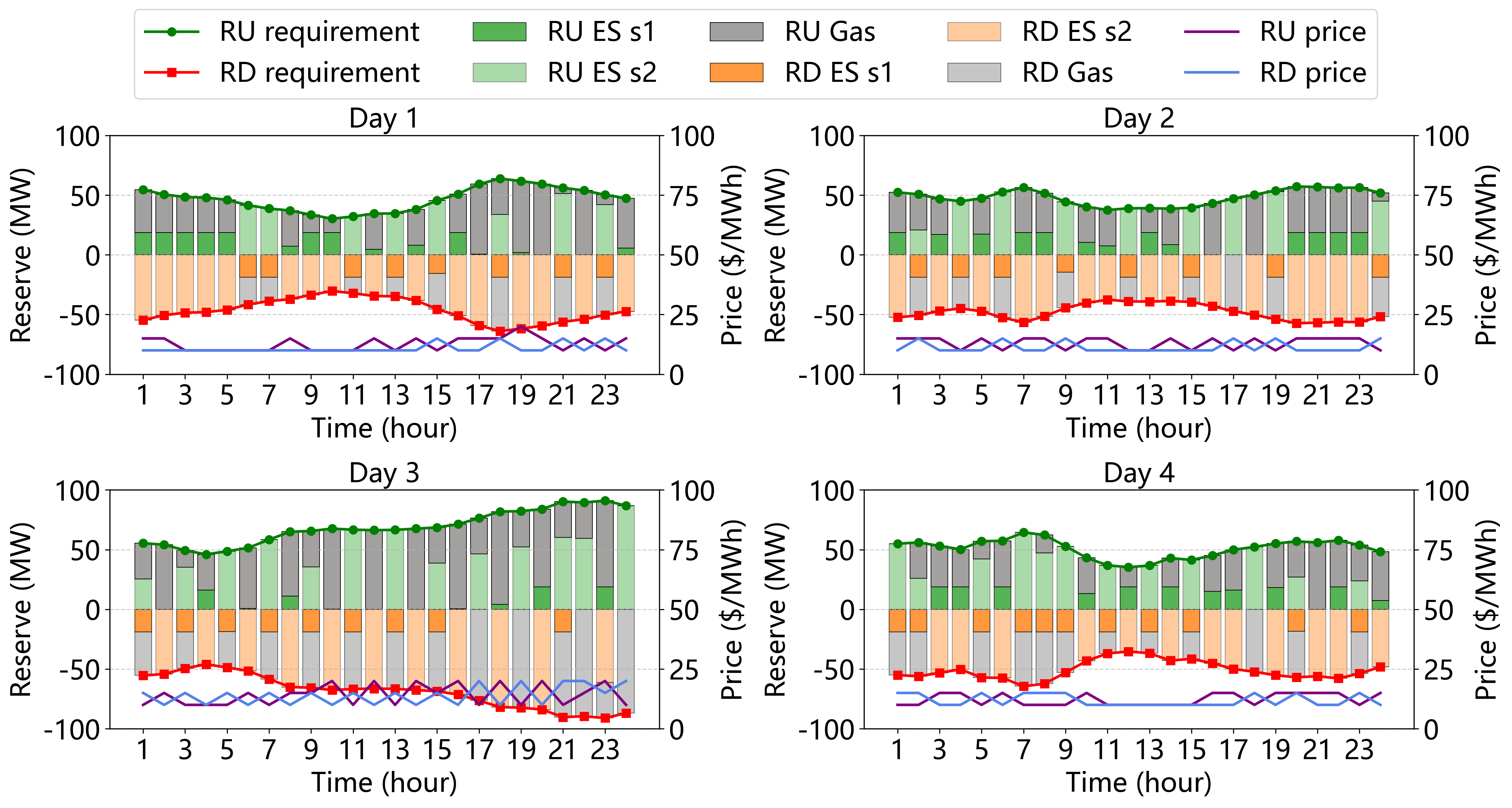}
    \caption{Reserve capacity balance of the joint day-ahead market.}
    \label{fig-da-reserve}
\end{figure}

In Figure \ref{fig-da-reserve}, while both traditional gas generators and the ESO contribute to satisfying the system's dynamic reserve requirements, the ESO exhibits a higher proportional market participation in the reserve balance compared to its role in the energy market. During specific trading periods, the allocated capacity of the ESO is sufficient to fulfill the system's reserve requirements. By frequently acting as the marginal provider, the ESO's block offers directly impact the reserve clearing prices.

\begin{figure}[!t]
    \centering
    \includegraphics[width=1\linewidth]{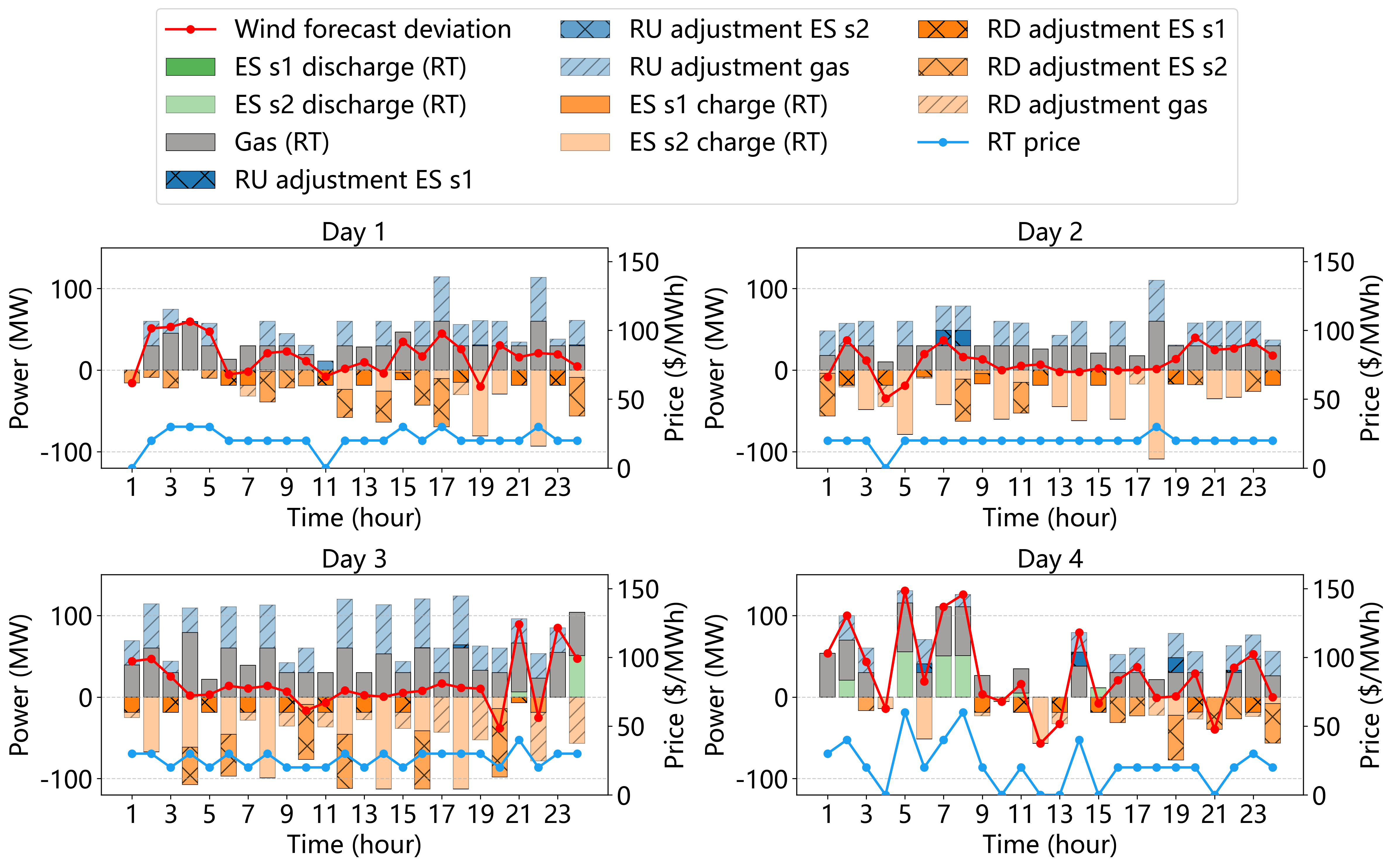}
    \caption{Balance of the real-time market.}
    \label{fig-rt}
\end{figure}

Real-time market dynamics are driven by the need to offset wind forecast deviations. Figure \ref{fig-rt} shows that the SO dispatches the deployment of day-ahead cleared reserves from both gas units and storage systems in Figure \ref{fig-da-reserve} to maintain physical balance in the real-time market. The ESO proves highly responsive in this context, executing frequent state transitions to absorb excess renewable generation or cover supply deficits. This cycling behavior underscores the value of maintaining flexible energy states through the explicit coupling of day-ahead and real-time market decisions.
\subsection{The Bidding Behavior of the ESO}

\begin{figure}[!t]
    \centering
    \includegraphics[width=1\linewidth]{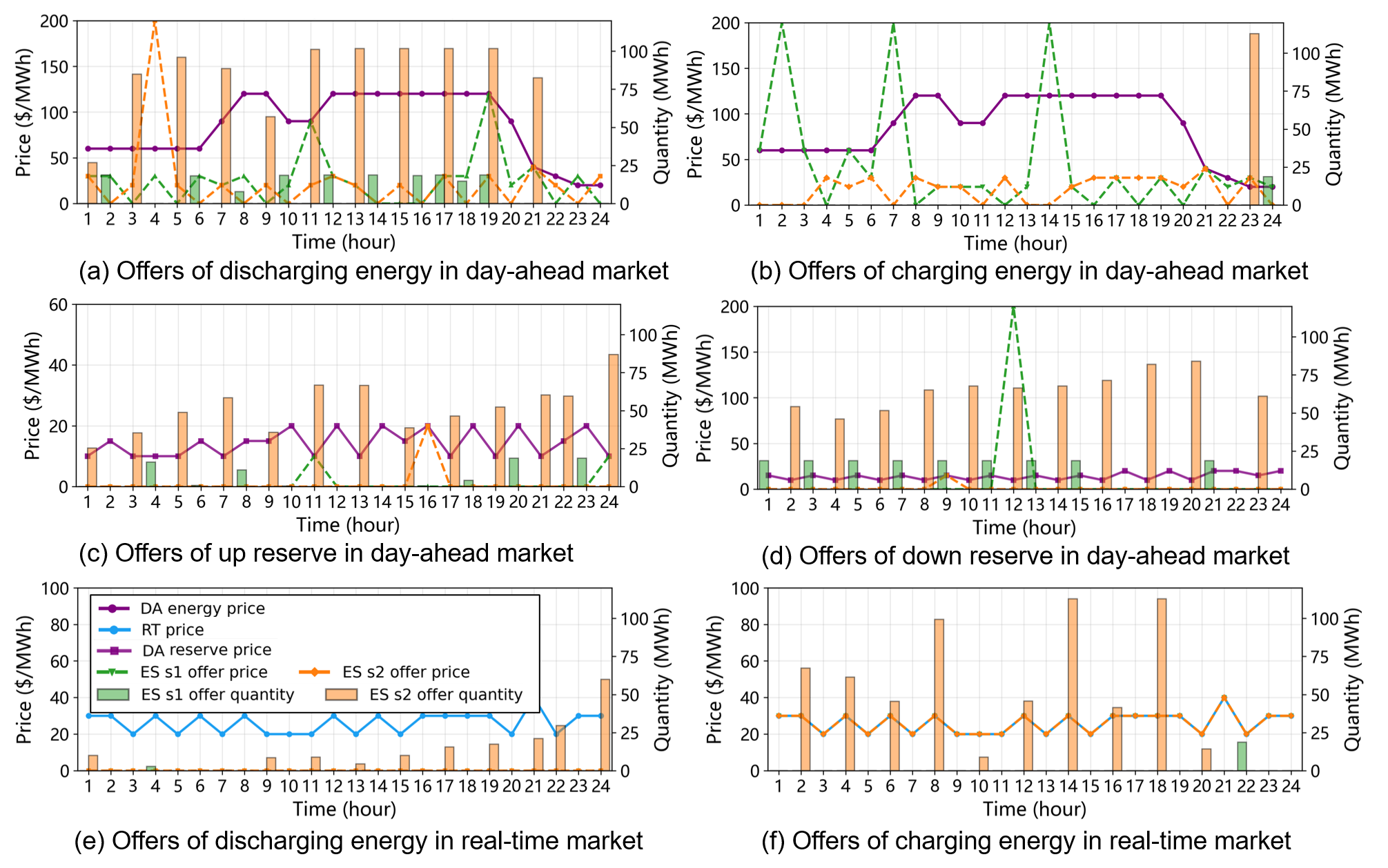}
    \caption{Offers of the ESO in multiple markets.}
    \label{fig-offer}
\end{figure}

Figure \ref{fig-offer} presents the optimal dispatch and market participation of the ESO over a typical day. The offer quantities of $s_1$ (green) and $s_2$ (orange) are shown on the left and right sides of each coordinate point, respectively. The two systems operate in a differentiated manner, reflecting their distinct technical characteristics and economic roles. In terms of optimal offer quantities, $s_1$, with a lower power-to-capacity ratio, primarily participates in the day-ahead market, whereas $s_2$, with a higher ratio, actively engages in both day-ahead and real-time markets, accompanied by larger energy fluctuations. Regarding offer prices, in most trading periods, both storage systems offer at prices less than or equal to the final clearing price, ensuring priority dispatch through the merit-order effect.

The temporal pattern of charging and discharging demonstrates that the ESO actively responds to both day-ahead and real-time price signals. It accumulates energy during low day-ahead price hours and discharges when prices are high or reserve deployment becomes economically attractive, confirming a strategic allocation of capacity between energy and reserve markets rather than full commitment to a single market. The selective real-time activation of day-ahead reserve commitments further illustrates the practical effect of coupling constraints between day-ahead cleared reserves and real-time deployment. Economically, Figure \ref{fig-offer} implies that the optimal capacity and operation results in balanced utilization of both storage systems, with neither remaining idle for extended periods and their operational profiles complementing each other across varying market conditions.

\begin{figure}[!t]
    \centering
    \includegraphics[width=1\linewidth]{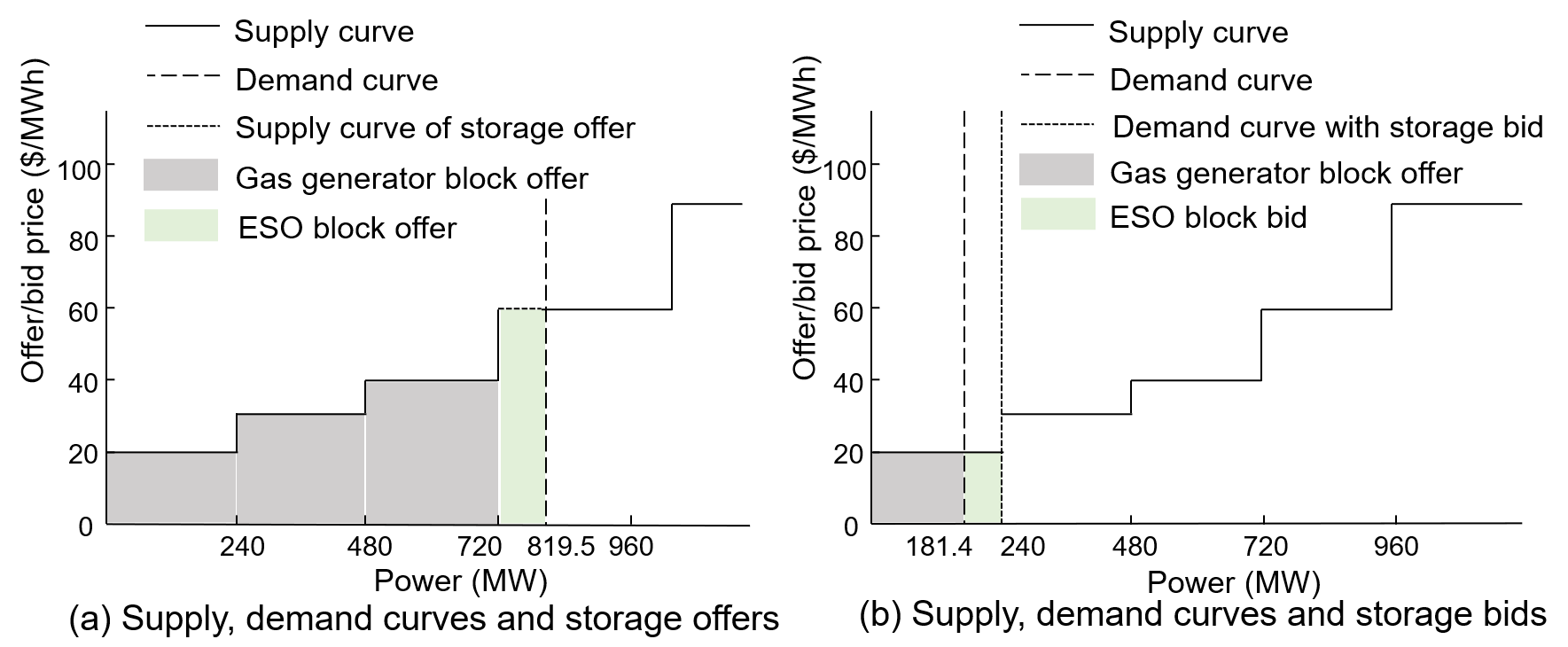}
    \caption{Supply, demand curves and storage offers/bids in the day-ahead market.}
    \label{fig:strategic}
\end{figure}

Figure \ref{fig:strategic} illustrates the shift in market equilibrium induced by the discharging and charging behaviors of the ESO's $s_2$ system at specific hours when $s_1$ remains idle in the day-ahead market. Note that the horizontal axis represents the residual power after accounting for priority-dispatched wind generation, thus originating at 0 MW. Figure \ref{fig:strategic}a demonstrates the market impact during the discharging phase. By submitting an offer block of 99.5 MW at \$60/MWh, the ESO inserts its capacity into the merit order. This strategic insertion effectively captures the market share that would otherwise belong to the fourth gas generator, allowing the ESO to maximize its dispatched volume without triggering a lower clearing price tier. Figure \ref{fig:strategic}b captures the ESO's price-maker behavior during the charging phase. The ESO's charging load shifts the system demand curve to the right. Here, the ESO strategically submits a bid block of 58.6 MW at \$20/MWh. This quantity is precisely calibrated to exhaust the remaining available capacity of the first gas generator (which is capped at 240 MW). If the ESO were to bid a volume greater than 58.6 MW with a price exceeding \$30/MWh, the market clearing mechanism would be forced to dispatch the second gas generator. This would elevate the marginal day-ahead energy price to \$30/MWh, subsequently increasing the ESO's overall charging costs.

\subsection{Profit Distribution}

The profit distribution detailed in Table \ref{tab:profit_distribution} quantifies the financial performance of the HESS across the four typical days. A distinct structural dependency emerges between the market layers. The day-ahead market consistently functions as the primary revenue stream for both storage systems. In contrast, the real-time market profit remains negative in nearly all modeled scenarios. The main reason behind this is that the clearing price in the real-time market is relatively lower than that in the day-ahead market. Therefore, the ESO tends to charge in the real-time market, which corresponds to Figure \ref{fig-da-energy} and \ref{fig-rt}.

\begin{table*}[!t]
\centering
\caption{Profit distribution of the ESO across typical days (\$)}
\label{tab:profit_distribution}
\footnotesize
\begin{tabular}{
    c c rrrrrr
}
\hline
{Day} & {\shortstack{Storage\\system}} 
& {\shortstack{Day-ahead\\ energy}} 
& {\shortstack{Day-ahead\\reserve}}  
& {\shortstack{Day-ahead\\ total profit}}  
& {\shortstack{Real-time\\ balance}}  
& {\shortstack{Total unweighted\\ profit}} 
& {\shortstack{Total weighted\\ profit}} \\
\hline
1 & $s_{1}$ &  6583.85 &  4039.78 &  10623.63 &  -2423.80 &  8199.82 & 2049.96 \\
1 & $s_{2}$ & 28032.06 & 10909.86 &  38941.92 & -14243.25 & 24698.67 & 6174.67 \\
2 & $s_{1}$ &  5272.93 &  4977.67 &  10250.60 &  -1601.34 &  8649.27 & 2162.32 \\
2 & $s_{2}$ & 32682.63 & 11266.96 &  43949.59 & -16878.93 & 27070.66 & 6767.66 \\
3 & $s_{1}$ & 17402.45 &  3813.10 &  21215.55 &  -3708.48 & 17507.06 & 4376.77 \\
3 & $s_{2}$ & 86191.36 & 14949.98 & 101141.34 & -26345.31 & 74796.03 & 18699.01\\
4 & $s_{1}$ &  3134.27 &  5366.63 &   8500.90 &   -364.14 &  8136.76 & 2034.19 \\
4 & $s_{2}$ & 11537.99 & 11140.89 &  22678.88 &   3670.70 & 26349.57 & 6587.39 \\
\hline
\end{tabular}
\end{table*}

This financial outcome aligns with the frequent real-time state transitions observed in the system balancing analysis (Figure \ref{fig-rt}). At the individual system level, performance exhibits a strong asymmetry driven by each technology’s physical parameters. The second system ($s_{2}$), characterized by a high power-to-capacity ratio, captures the majority of joint operational profit, particularly in the day-ahead energy market, where it executes rapid, high-volume transfers during volatile pricing windows. For instance, on the first typical day, $s_{2}$ generates nearly four times the unweighted total profit of the first system ($s_{1}$). Conversely, $s_{1}$’s revenue shifts toward the day-ahead reserve market. On the fourth typical day, its reserve revenue explicitly exceeds its energy revenue. Owing to its low power-to-capacity ratio, the optimization model preferentially allocates $s_{1}$ to provide baseline reserve margins, freeing $s_{2}$ for aggressive arbitrage cycling. These numerical outcomes corroborate the differentiated bidding behaviors identified earlier (Figure \ref{fig-offer}) and validate the modeling necessity of treating heterogeneous storage systems with independent decision variables. Table \ref{tab:profit_distribution_vali} presents the profit distribution across out-of-sample days, which aligns with the typical day analysis. 

\begin{table}[!t] 
\centering
\caption{Profit distribution of the HESS across out-of-sample days (\$)}
\label{tab:profit_distribution_vali}
\footnotesize
\begin{tabular}{ccrrr}
\hline
Day & \shortstack{Storage\\system} & 
\shortstack{Day-ahead\\ energy} & 
\shortstack{Day-ahead\\ reserve} &
\shortstack{Real-time\\ balance}\\
\hline
5 & $s_{1}$ & 3144.58 & 5254.65 & -1199.99 \\
5 & $s_{2}$ & 9067.39 & 13185.95 & 3954.44 \\
6 & $s_{1}$ & 5370.95 & 4876.74 & -2364.68 \\
6 & $s_{2}$ & 28366.46 & 10828.00 & -13203.52 \\
7 & $s_{1}$ & 9161.76 & 3860.67 & -3716.51 \\
7 & $s_{2}$ & 38292.17 & 10999.23 & -17063.77 \\
8 & $s_{1}$ & 1881.94 & 6158.27 & -1314.35 \\
8 & $s_{2}$ & 14250.49 & 12900.26 & -5402.03 \\
9 & $s_{1}$ & 13861.38 & 4652.40 & -4872.98 \\
9 & $s_{2}$ & 71145.96 & 14049.46 & -27481.86 \\
10 & $s_{1}$ & 11148.12 & 3594.57 & -3957.56 \\
10 & $s_{2}$ & 45818.55 & 10643.38 & -20895.32 \\
11 & $s_{1}$ & 1977.92 & 5910.18 & -1596.25 \\
11 & $s_{2}$ & 13305.24 & 12667.13 & -2233.28 \\
12 & $s_{1}$ & 2178.10 & 5447.20 & -1815.91 \\
12 & $s_{2}$ & 8943.31 & 12510.12 & 4125.34 \\
\hline
\end{tabular}
\end{table}

\subsection{Internal Transmission between Two Storage Systems}

The proposed framework explicitly models power flows among facilities, though such internal transmission did not occur in prior experiments. Here we interpret internal power transmission between two storage systems under a co-located deployment with limited grid access. The two systems, while jointly owned and operated by a single ESO, are assumed to share a common point of connection with other on-site resources (e.g., renewable generators or additional storage assets). Consequently, available grid interconnection capacity becomes a binding operational constraint that restricts simultaneous charging or discharging of individual systems. Under these conditions, internal power transmission functions not as an arbitrage mechanism but as a coordination strategy enabling the operator to reallocate power within the hybrid system when direct grid access is constrained.

Table \ref{tab:investment_scenarios} shows that internal power transmission is activated primarily when grid charging or discharging of one or both storage systems is constrained, indicating that it arises from physical limitations rather than price signals. Meanwhile, capacity planning decisions are highly sensitive to grid access, indicating that the economic value of the hybrid configuration depends on unrestricted simultaneous coordination of both systems, and that access limitations critically affect the necessity of dual-system investment.

\begin{table*}[!t] 
\centering
\caption{Capacity planning and operational outcomes under different grid access constraint scenarios}
\label{tab:investment_scenarios}
\small 
\begin{tabular}{l rrr c}
\hline
Grid access constraint scenario 
& {\shortstack{$E_{s_1}$\\ (MWh)}} 
& {\shortstack{$E_{s_2}$\\(MWh)}} 
& {\shortstack{Profit \\(\$/day)}} 
& {\shortstack{Internal power \\reallocation required}}\\
\hline
Shared grid import capacity limit at the co-located site &  0.63 &  98.00 & 22086.72 & Yes \\
Grid import congestion affecting $s_1$                   &  0.02 & 117.67 & 23662.22 & Yes \\
Grid import congestion affecting $s_2$                   & 44.14 & 113.32 & 23965.39 & No  \\
Shared grid export capacity limit at the co-located site &  0.14 & 120.48 & 18497.63 & Yes \\
Grid export congestion affecting $s_1$                   &  0.14 & 108.14 & 23565.29 & Yes \\
Grid export congestion affecting $s_2$                   & 99.57 & 104.00 & 20078.92 & Yes \\
\hline
\end{tabular}
\end{table*}

\section{Conclusion}

This paper proposes a bi-level optimization framework to determine the optimal capacity planning and multi-market participation strategy for an independent ESO managing a HESS. The model represents the operator as a price-maker in both the day-ahead joint energy and reserve market and the real-time balancing market. The upper-level problem maximizes the profit of the operator by co-optimizing the capacities of two heterogeneous storage systems and their respective price-quantity offers. The lower-level problems simulate the market clearing processes. The formulation is transformed into a single-level MPEC and solved via Benders' decomposition.

Numerical results demonstrate that the proposed framework effectively captures the market influence of large-scale storage and maximizes overall profitability. The simulation outcomes reveal that preserving the heterogeneity enables the development of differentiated bidding strategies. The storage system characterized by a high power-to-capacity ratio actively participates in both day-ahead and real-time energy markets to perform aggressive arbitrage, capturing the majority of the operational profit. In contrast, the system with a low power-to-capacity ratio is preferentially allocated to provide baseline reserve margins in the day-ahead market. Internal power transmissions between the distinct systems provide critical flexibility to mitigate grid connection limits and directly affect the optimal capacity planning decisions.

Future research will investigate the impact of uncertainties regarding renewable generation and load forecasts on the planning and operation of the HESS. Expanding the framework to include detailed degradation models for distinct storage technologies and analyzing the market equilibrium among multiple storage operators constitute other valuable directions for future work.

\bibliographystyle{IEEEtran}
\bibliography{ref.bib}

\end{document}